\documentclass[twocolumn, showpacs,preprintnumbers,amsmath,amssymb]{revtex4}

\usepackage{verbatim}
\usepackage{amsmath}
\usepackage[pdftex]{graphicx}

\begin{document}

\addtolength{\textheight}{1.2cm}
\addtolength{\topmargin}{-0.5cm}

\newcommand{\etal} {{\it et al.}}

\title{On the size of linear superpositions in molecular nanomagnets}

\author{F. Troiani}

\affiliation{S3 Istituto Nanoscienze-CNR, Modena, Italy}

\author{P. Zanardi}

\affiliation{
Department of Physics and Astronomy, and Center for Quantum Information Science \& Technology, University of Southern California, Los Angeles, CA 90089-0484}
\affiliation{    Centre for Quantum Technologies,  National University of Singapore,  2 Science Drive 3, Singapore 117542}

\date{\today}

\begin{abstract}

Molecular nanomagnets are zero-dimensional spin systems, that exhibit quantum mechanical behavior at low temperatures. Exploiting quantum-information theoretic measures, we quantify here the size of linear superpositions that can be generated within the ground multiplet of high- and low-spin nanomagnets. Amongst the former class of systems, we mainly focus on Mn$_{12}$ and Fe$_8$. General criteria for further increasing such sizes are also outlined, and illustrated in the case of spin rings. The actual character (micro or macroscopic) of linear superpositions in low-spin systems is inherently ill-defined. Here, this issue is addressed with specific reference to the Cr$_7$Ni and V$_{15}$ molecules, characterized by an $S=1/2$ ground state. In both cases, the measures we obtain  are larger than those of a single $s=1/2$ spin, but not proportionate to the number and value of the constituent spins.

\end{abstract}


\maketitle

Molecular nanomagnets represent prototypical examples of engineerable quantum systems. 
In fact, their physical properties can be widely tuned by chemical synthesis, and quantum coherence effects show up in their spin dynamics \cite{Gatteschi,Furrer13}. 
These effects include quantum tunneling of the molecule spin, resulting in a speed-up of the magnetization relaxation \cite{Thomas96,Friedman96}, and quantum phase interference \cite{Wernsdorfer99}. 
Besides, microwave-induced quantum coherences \cite{Ardavan07,Takahashi09} and Rabi oscillations \cite{Bertaina08,Schlegel08} were recently observed in a wide class of spin clusters. 
These experimental achievements, along with the microscopic understanding and chemical control of the decoherence processes \cite{Troiani08,Wedge12}, will possibly enable the use of molecular nanomagnets for quantum-information processing \cite{Leuenberger01,Troiani05a}. 

Amongst molecular nanomagnets with dominant exchange interaction, a prominent distinction is that between high- and low-spin systems. In the former class of spin clusters, the ground $S$ multiplet may include classical-like states that are macroscopically different from one another, and whose linear superpositions can thus be regarded as Schr\"odinger cat states \cite{Schroedinger35}. In the latter systems, antiferromagnetic interactions result instead in ground states with low $S$ and highly non-classical features, such as quantum entanglement \cite{Siloi12} or N\'eel-vector tunneling \cite{Chiolero98,Santini05,Waldmann09}. Hereafter, we theoretically investigate the size of linear superpositions that can be - or have already been - generated in both these kinds of molecular nanomagnets. In the case of high-spin molecules, we quantify the actual macroscopicity of linear superpositions. In other words, we determine to which extent their sizes are proportionate to the number and value of the constituent spins, and thus approach the theoretical maxima. We initially focus on the most celebrated single-molecule magnets, [Mn$_{12}$O$_{12}$(CH$_3$COO)$_{16}$(H$_2$O)$_4$] \cite{Lis80} and [Fe$_8$O$_2$(OH)$_{12}$(tacn)$_6$]Br$_8$(H$_2$O)$_9$ (tacn=1,4,7-triazcyclononane) \cite{Wieghardt84}. In both these nanomagnets (hereafter referred to as Mn$_{12}$ and Fe$_8$) the ground state is characterized by a ferrimagnetic ordering, with the total spin $S$ resulting from the inequivalence of the two sublattices \cite{Gatteschi}. The role played by such inequivalence is further discussed in reference to a prototypical class of spin rings, which includes the [Mn(hfac)$_2$(NITPh)]$_6$ molecule (Mn$_6$) \cite{Caneschi88}, 
and a general criterion for maximizing the size of linear superpositions is outlined. 
In the case of low-spin molecules, the actual character of linear superpositions seems intrinsically ill-defined. In fact, these might appear to be either micro or macroscopic, depending on whether one refers to the total spin of the ground state or to the number of spins that form the cluster. Here, we quantify the size of linear superpositions that have been recently generated in two significantly different prototypes of $S=1/2$ molecular nanomagnets, namely [Cr$_7$NiF$_8$\{O$_2$CC(CH$_3$)$_3$\}$_{16}$] \cite{Ardavan07} and [V$_{15}^{IV}$As$_6$O$_{42}$(H$_2$O)] \cite{Bertaina08} molecules, hereafter referred to as Cr$_7$Ni and V$_{15}$. In both cases, the measures we obtain are significantly larger than those of a single $s=1/2$ spin, but not proportionate to the number and value of the constituent spins.

The problem of quantifying the size of a linear superposition has been addressed from different perspectives, mainly to assess the actual macroscopicity of quantum coherences in infinite physical systems \cite{Leggett80,Dur02,Shimizu02,Bjork04,Marquardt08,Frowis12}. 
There, the question whether or not a linear superposition is macroscopic is answered by considering the limit of different measures as the number $N$ of the microscopic subsystems tends to infinity. In the present paper, we use two of such measures to quantify the size of linear superpositions in zero-dimensional spin systems. The first measure is based on the quantum Fisher information \cite{Frowis12}, and quantifies both  the non-classicality of the linear superposition $ | \Psi \rangle = ( | \Psi_1 \rangle + | \Psi_2 \rangle ) / \sqrt{2} $ and the classical-like character of its components in terms of the respective quantum fluctuations of single-spin operators. As showed in the following, the advantage of such measure in the present context is at least twofold. Firstly, it quantifies the distinguishability between the two superposed states by means of a physically intuitive quantity, such as the spin projection of the constituent ions. Secondly, it can be expressed in terms of spin-pair correlation functions, that are now experimentally accessible in molecular spin clusters \cite{Baker12}. The second measure we consider quantifies the size of a linear superposition in terms of the possibility to distinguish between the two components $ | \Psi_1 \rangle $ and $ | \Psi_2 \rangle $ by means of local measurements. The size of $ | \Psi \rangle $ is thus identified with the number of subsystems into which the spin cluster can be partitioned, such that the which-component information is available in each of the subsystems \cite{Korsbakken07}. The notion of distinguishability can be itself translated in terms of measurement outcomes, but in a way that makes it of little experimental relevance in the present context. In fact, 
the projective measurements of the individual spins (or of arbitrary subsets of them) that would be required to estimate the distinguishability are presently unfeasible in molecular nanomagnets. On the other hand, this measure has the advantage of being independent on the specific class of one-body operators that we use with the Fisher information, and is thus suitable for comparing the size of linear superpositions obtained in completely different systems.

The paper is organized as follows. In Section I we introduce the measures that are used to quantify the size of the linear superpositions. In Section II we investigate spin clusters that allow the generation of large linear superpositions. In particular, we present the cases of Mn$_{12}$ and Fe$_8$, and discuss which general features of a spin cluster maximize the considered measures. Section III is devoted to nanomagnets with low-spin and highly frustrated ground states, and specifically to Cr$_7$Ni and V$_{15}$. Finally, we summarize our findings and draw the conclusions in Section IV. 

\section{Size of linear superpositions in spin clusters}

In the following, we consider linear superpositions of the form
\begin{equation}\label{eq01}
| \Psi \rangle = \frac{1}{\sqrt{2}} ( | \Psi_1 \rangle + | \Psi_2 \rangle ) ,
\end{equation}
whose components are eigenstates of the spin Hamiltonian $H_0$ of the molecular nanomagnet. For the systems of interest, $H_0$ commutes with both ${\bf S}^2$ and $S_z$, and its dominant term ($H_{Exc}$) includes the exchange couplings between the $N$ spins. The considered states $ | \Psi_1 \rangle $ and $ | \Psi_2 \rangle $ belong to the ground $S$-multiplet, and are characterized by defined values of $S_z$ ($M_1$ and $M_2$, respectively). 

\subsection{Quantum Fisher information}

The first criterion we use for quantifying the size of the linear superposition is based on the quantum Fisher information \cite{Frowis12}. For pure states, such quantity coincides, up to a multiplicative constant, with the variance of the relevant operator $X$:
\begin{equation}
\mathcal{F} (\Psi) = 4 \max_{X} \left[ \langle \Psi | X^2 | \Psi \rangle - \langle \Psi | X | \Psi \rangle^2 \right] .
\end{equation}
Here, $ X = \sum_{i=1}^N \hat{\bf n}_i \cdot {\bf s}_i $ is a generic sum of single-spin components, with site-dependent orientations, that are specified by the versors $\hat{\bf n}_i$. The quantum Fisher information thus reads:
\begin{eqnarray} \label{explFI}
\mathcal{F} (\Psi)  & = & 4        \sum_{\alpha , \beta = x,y,z} \sum_{i,j=1}^N 
n_{i,\alpha} n_{j,\beta} \langle \Psi | s_{i,\alpha} s_{j,\beta} | \Psi \rangle 
\nonumber\\
                    & - & 4 \left( \sum_{\alpha         = x,y,z} \sum_{i  =1}^N 
n_{i,\alpha}             \langle \Psi | s_{i,\alpha}             | \Psi \rangle \right)^2 .
\end{eqnarray}
For any given $ | \Psi \rangle $, we compute the one- and two-spin expectation values that enter the above expression, and derive the set of versors ${\bf n}_i$ that maximize $\mathcal{F} (\Psi) $. At a qualitative level, the optimal operator $X$ should be a well defined quantity within the states $ | \Psi_1 \rangle $ and $ | \Psi_2 \rangle $, with very different expectation values in the two cases. Therefore, it efficiently distinguishes between the two components of the linear superposition. 

For a given spin cluster, the theoretical maximum of the Fisher information is given by: 
\begin{equation}\label{psimax}
\mathcal{F}_{max} = 
                4 \left( \sum_{i=1}^{N} \| \hat{\bf n}_i \cdot {\bf s}_{i} \| \right)^2 \!\!\!\!
              =\! 4 \!\left( \sum_{i=1}^{N} s_i \right)^2 \!
\!,
\end{equation}
where the norm of the spin operator is defined as its largest eigenvalue (in modulus).
This value corresponds to the state 
$
|\Psi_{max}\rangle = (|\Psi_1^{max}\rangle+|\Psi_2^{max}\rangle)/\sqrt{2}
$,
whose components read
\begin{equation}\label{xxx}
|\Psi_{k=1,2}^{max}\rangle = \otimes_{i=1}^N | \hat{\bf n}_i \cdot {\bf s}_i = (-1)^{k+1} s_i \rangle .
\end{equation}
These states are characterized by opposite expectation values 
($ \langle \Psi_k^{max} | X | \Psi^{max}_k \rangle = (-1)^{k+1} \sum_{i=1}^N s_i $) 
and vanishing fluctuations of $X$.

In most of the cases considered hereafter, the operator $X$ that maximizes $ \mathcal{F} (\Psi) $ coincides with the staggered magnetization $ S_z^* $:
\begin{equation}
X = S_z^* = S_z^A - S_z^B = \sum_{k=1}^{N_A} s^A_{k,z} - \sum_{l=1}^{N_B} s^B_{l,z} ,
\end{equation}
where $A$ and $B$ are two sublattices into which the spin cluster is partitioned ($N_A+N_B=N$). Each sublattice includes all the spins $s_i$ with the same sign of the expectation value $ \langle \Psi_k | s_{z,i} | \Psi_k \rangle $, for any given component $k=1,2$ of the linear superposition. 
The components of $|\Psi_{max}\rangle$ then take the form of collinear spin states:
\begin{equation}\label{wfmax}
|\Psi_{k}^{max}\rangle =  
   \otimes_{i=1}^{N_A} |m_{i}^A=\pm s_i^A\rangle 
   \otimes_{j=1}^{N_B} |m_{j}^B=\mp s_j^B\rangle ,
\end{equation}
where the first and second signs correspond to $k=1$ and $k=2$, respectively.

The quantum Fisher information can be used to quantify the size of the linear superposition $ | \Psi \rangle $ \cite{Frowis12}: 
\begin{equation}
D_{FI} (\Psi) = \frac{\mathcal{F}( \Psi )}{4 \sum_{i=1}^{N} s_i}. 
\end{equation}
Given the above normalization, $ D_{FI} (\Psi) $ can take any value in the interval $ [ 0 , \sum_{i=1}^{N} s_i ] $. The maximum size of a linear superposition that can in principle be generated in a given spin cluster thus depends not only on the number $N$ of the constituent spins, but also on their values $s_i$. 
The restriction to operators $X$ that are linear combinations of single-spin terms allows us to relate the size $ D_{FI} (\Psi) $ to experimentally accessible quantities, such as spin-pair correlation functions \cite{Baker12}.

The absolute value of the Fisher information quantifies the size of the linear superposition in terms of the quantum fluctuations of the operator $X$. However, it doesn't discriminate between the contribution to such fluctuations coming from the components $| \Psi_1\rangle$ and $| \Psi_2\rangle$, and the one that results from their linear superposition. In a Schr\"odinger-cat state, for example, the latter contribution should dominate on the former one \cite{Schroedinger35}: the superposed states should in fact approach a classical character, and be characterized by the smallest amount of quantum fluctuations allowed by the uncertainty relations. In this perspective, a suitable figure of merit for the linear superposition is represented by the relative Fisher information, where the variance of the relevant operator in $ | \Psi \rangle $ is normalized to the average variance in $ | \Psi_1 \rangle $ and $ | \Psi_2 \rangle $ \cite{Frowis12}: 
\begin{equation}
D_{RFI} (\Psi) 
= 
\frac{\mathcal{F}_\Psi (\Psi)
}{\frac{1}{2}
\left[\mathcal{F} (\Psi_{1}) + \mathcal{F}( \Psi_{2} )\right]} .
\end{equation}
Here, all the functions $ \mathcal{F}$ refer to the operator $X$ that maximizes $D_{FI} (\Psi)$. The measure $ D_{RFI} (\Psi) $ is not bounded from above, and diverges if the components of the linear superpositions are eigenstates of $X$.

\begin{figure}[ptb]
\begin{center}
\includegraphics[width=8.0cm]{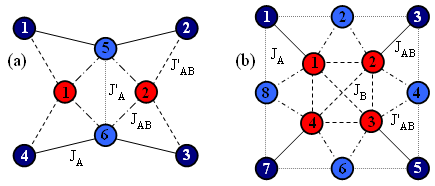}
\end{center}
\caption{(color online) (a) Spin cluster corresponding to the Fe$_{8}$ molecule. The eight $s=5/2$ spins are divided into the two sublattices $A$ (light and dark blue) and $B$ (red), defined by their relative orientation in the system ground state. Equivalent spins are denoted by the same color. Same line style corresponds to identical values of the exchange coupling between neighboring spin pairs. (b) Spin cluster corresponding to the Mn$_{12}$ molecule, where $s_i^A=2$ and $s_j^B=3/2$, with the same conventions as above. 
}
\label{Fe8Mn12}
\end{figure}

\subsection{Distinguishability by local measurements}

The size of the linear superposition $ |\Psi\rangle $ can also be quantified in terms of the maximal number of subsystems that carry the which-component information \cite{Korsbakken07}. The maximum probability of successfully discriminating between $ | \Psi_1\rangle $ and $ | \Psi_2 \rangle $ through a measurement on the subsystem $A_l$ of the spin cluster is given by:
\begin{equation}
P_l = \frac{1}{2} + \frac{1}{4} \| \rho^{(1)}_{l} - \rho^{(2)}_{l} \|_1 ,
\end{equation}
where $ \rho^{(k)}_{l} $ is the reduced density matrix of $A_l$ obtained from the state $ | \Psi_k \rangle $, and $\| X \|_1 := \sum_i |\lambda_i| $ (with $\lambda_i$ the eigenvalues of $X$) is the trace norm. 
The size of $ | \Psi \rangle $ based on the distinguishability between the two components by local measurements is given by: 
\begin{equation} \label{eqlm}
D_{LM} (\Psi , \delta) = \max_{\{A_l\}} n ,
\end{equation} 
where $n$ is the number of subsystems $A_l$ into which the $N$-spin cluster is partitioned. The maximization is performed over all the partitions such that $ P_l > 1 - \delta $ for all $l=1, \dots , n$, where $\bigcup_{l=1}^n A_l = \mathcal{S}$ and $\mathcal{S}$ is the $N$-spin system. Given a spin cluster, the present measure achieves its theoretical maximum for a state such as $ | \Psi_{max} \rangle $ (Eq. \ref{xxx}), being
$ D_{LM} ( \Psi_{max} , \delta ) = N $ for any finite $\delta$. 

The size quantification based on the local distinguishability has the advantage, with respect to $D_{FI}$ and $D_{RFI}$, of being independent on the specific class of operators $X$. It thus allows the comparison between the sizes of linear superpositions that we obtain for the molecular nanomagnets with those estimated for different systems. On the other hand, the connection to experimentally accessible quantities is less straightforward. The maximum probability $P_l$ is in fact obtained by a projective measurement in the basis that diagonalizes $ \rho^{(1)}_{l} - \rho^{(2)}_{l} $ (see Ref. \cite{Korsbakken07} and references therein). In the case where $A_l$ includes more than one spin, the expression of the relevant observable might be a non-trivial function of the spin operators. Besides, projective measurements of single spins, or of specific subsystems of the spin cluster, are presently unavailable in molecular magnetism.

\section{High-spin molecules}\label{SecHSM}

The molecular nanomagnets where one can in principle generate the largest linear superpositions are those characterized by a ground state with a large spin $S$. In particular, the spin-polarized ground states ($M_1=+S$ and $M_2=-S$) are expected to have a classical-like character, and to be highly distinguishable in terms of single-spin operators, as should be the case with the components of a Schr\"odinger cat state. In the first part of the present Section, such expectations are tested for the prototypical high-spin nanomagnets, namely Mn$_{12}$ and Fe$_8$  (Subsec. \ref{subsec1}) \cite{Gatteschi}. In order to put in a wider perspective the results concerning these two molecules, we discuss how the sizes of linear superpositions can be possibly increased by suitably modifying the exchange couplings or the geometry of the spin cluster (Subsec. \ref{subsec3}). 

\subsection{Prototypical high-spin nanomagnets} \label{subsec1}

The dominant part $H_{Exc}$ in the spin Hamiltonian of Mn$_{12}$ and Fe$_8$ corresponds to the isotropic exchange interaction between neighboring spins. The total spin $S$ can thus be regarded as a good quantum number. Smaller terms, such as those related to crystal field, can be treated at a perturbative level and projected within the ground $S$-multiplet \cite{Gatteschi}. In particular, the crystal field anisotropy is often assumed to be quadratic in the total spin operators, and its main term reflects an axial symmetry: $H_{ACF}=DS_z^2$ (with $D<0$), where $z$ is the principal symmetry axis of the molecule. $H_{ACF}$ removes the degeneracy within the ground $S$ multiplet of $H_{Exc}$, giving rise to the characteristic double-well potential, with the two degenerate ground states $M= \pm S$. Hereafter we consider linear superpositions of the form Eq. \ref{eq01}, where $|\Psi_1\rangle$ and $|\Psi_2\rangle$ are eigenstates of $ H_0 = H_{Exc} + H_{ACF} $. Coherences between such eigenstates can be induced by additional and smaller terms in the spin Hamiltonian $H$ that don't commute with $S_z$, such as the rhombic crystal field $ H_{RCF} = E (S_x^2-S_y^2) $, or by a transverse magnetic field. Alternatively, arbitrary linear superpositions can be generated by multifrequency pulse sequences, that exploit the removed degeneracy between the different $ \Delta M = \pm 1 $ transitions \cite{Leuenberger01}.

\subsubsection{The Mn$_{12}$ molecule} 

The magnetic core of the Mn$_{12}$ molecule essentially consists of an external ring, formed by eight Mn$^{3+}$ ions (each carrying an $s=2$ spin), and four internal Mn$^{4+}$ ions (with $s=3/2$). Based on their relative orientation, these spins can be grouped in two sublattices, labeled $A$ and $B$, that are formed by the external and internal spins, respectively (Fig. \ref{Fe8Mn12}(b)) \cite{Gatteschi}. The dominant part $H_{Exc}$ of the spin Hamiltonian includes the exchange interactions between neighboring spins belonging to the same sublattice ($H_A$ and $H_B$) or to different ones ($H_{AB}$):
\begin{eqnarray} 
\label{HamMn12a}
H_A\!\!    & = &\! J_A \sum_{i=1}^8                  {\bf s}^A_i \cdot {\bf s}^A_{i+1} , \ 
H_B\!      =\!   J_B \sum_{i=1}^4 \sum_{j=1}^{i-1} {\bf s}^B_i \cdot {\bf s}^B_j     
\\
\label{HamMn12b}
H_{AB}\!\! & = &\!\! J_{AB} \!\! \sum_{i=1}^4 {\bf s}^A_{2i-1}\!\! \cdot {\bf s}^B_{ i}\!\! +
            \! J_{AB}' \!\sum_{i=1}^4 {\bf s}^A_{2i  } \!\cdot\! ({\bf s}^B_{i}+{\bf s}^B_{i+1})
\end{eqnarray} 
where $s^A_i=2$, $s^B_j=3/2$ ($ {\bf s}_9^A \equiv {\bf s}_1^A $ and $ {\bf s}_5^B \equiv {\bf s}_1^B $). The attempts made to estimate the above exchange couplings have not led so far to unanimous conclusions \cite{Gatteschi,Furrer13}. Hereafter, we report the results obtained with two plausible but significantly different sets of parameters. The comparison between the two cases might provide some insight into the role played by competing exchange interactions and spin frustration (see also Subsect. \ref{subsec3}). 
The first set of exchange couplings we refer to is given by \cite{Raghu01}: 
$J_A=-64.5\,$K, $J_B=85\,$K, $J_{AB} = 215\,$K, $J_{AB}'=85\,$K. 
The ground-state multiplet of $H_{Exc} = H_A + H_B + H_{AB} $ belongs to the $S=10$ subspace. 
Taking as a reference the linear superposition between the ground states $M_1=10$ and $M_2=-10$, the operator that maximizes $ \mathcal{F} ( \Psi )$ (Eq. \ref{explFI}) is $ X = S_z^* $, corresponding to $ \hat{\bf n}_{i} = \hat{\bf z} $ ($ \hat{\bf n}_{i} = -\hat{\bf z} $) for the spins belonging to sublattice $A$ ($B$).

We start by quantifying the classical-like character of the maximally polarized ground state $ | M = 10 \rangle $. The classical ground state (Eq. \ref{psimax}) represents the main configuration in $ | M = 10 \rangle $, the overlap between the two being 
$ | \langle M=10 | \Psi_{1}^{max} \rangle | = 0.307 $. The average value and the quantum fluctuations of the staggered magnetization provide alternative means to quantify the resemblance between quantum and classical ground states. Here, the expectation value of $S_z^*$ is given by 
$ \langle M=10 | S_z^* | M=10 \rangle = 17.6 $, 
out of a maximum of $ \langle \Psi_{1}^{max} | S_z^* | \Psi_{1}^{max} \rangle = 22 $. 
The variance is instead given by 
$ \mathcal{V}_{M=10} (S_z^*) = 7.0 $
(where $\mathcal{V}_{\Phi} (X) = \langle\Phi | X^2 | \Phi\rangle - \langle\Phi | X | \Phi\rangle^2 $),
as opposed to the absence of fluctuations ($ \mathcal{V}_{\Psi_{1}^{max}} (S_z^*) = 0 $) that characterizes the classical ground state. Ground states with decreasing values of $M > 0$ are characterized by a decreasing resemblance to a classical state. In particular, the expectation value $ \langle M | S_z^* | M \rangle $ and the variance $ \mathcal{V}_{M} $ undergo respectively a linear decrease and an exponential increase as $M$ varies from 10 to 1.

\begin{figure}[ptb]
\begin{center}
\includegraphics[width=8.0cm]{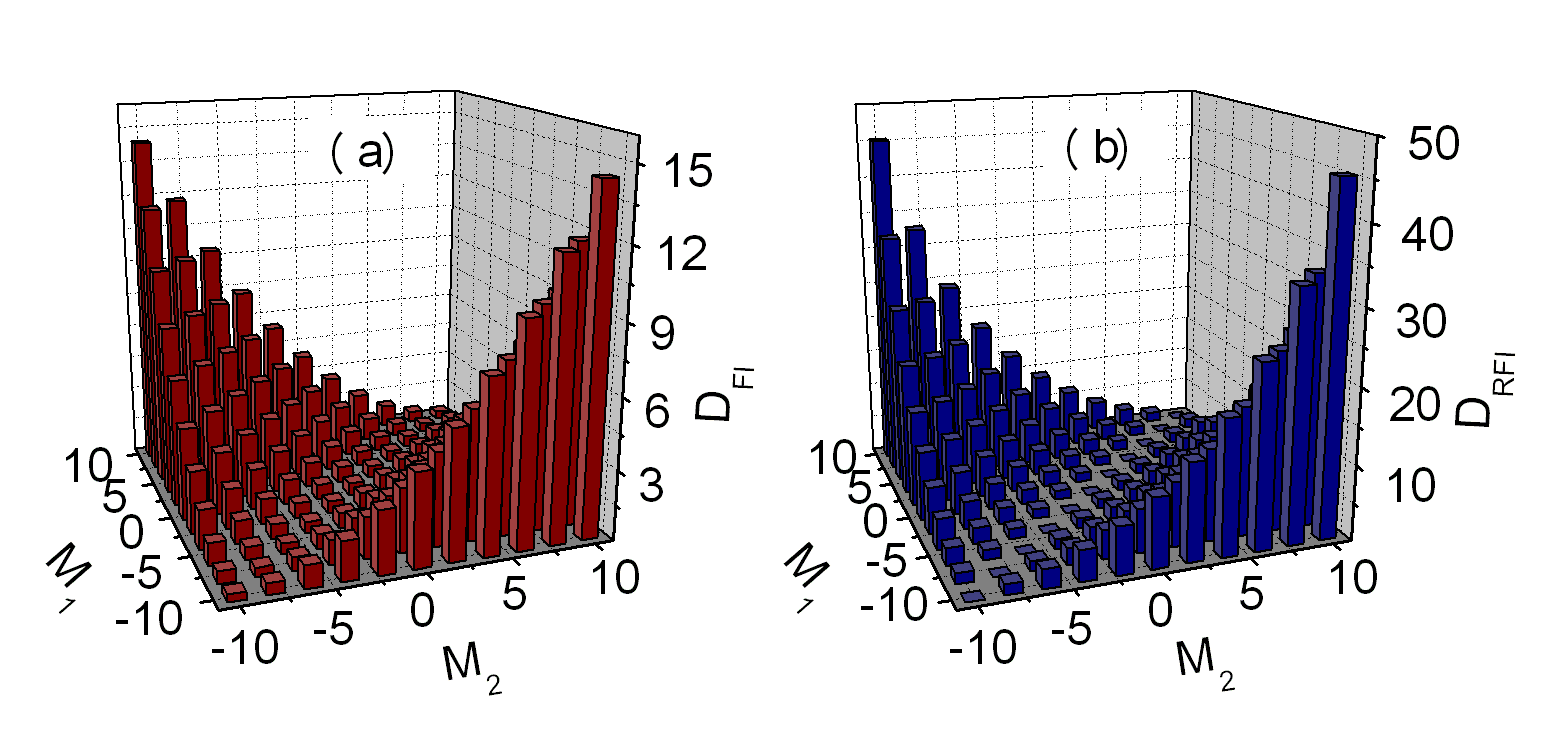}
\includegraphics[width=8.0cm]{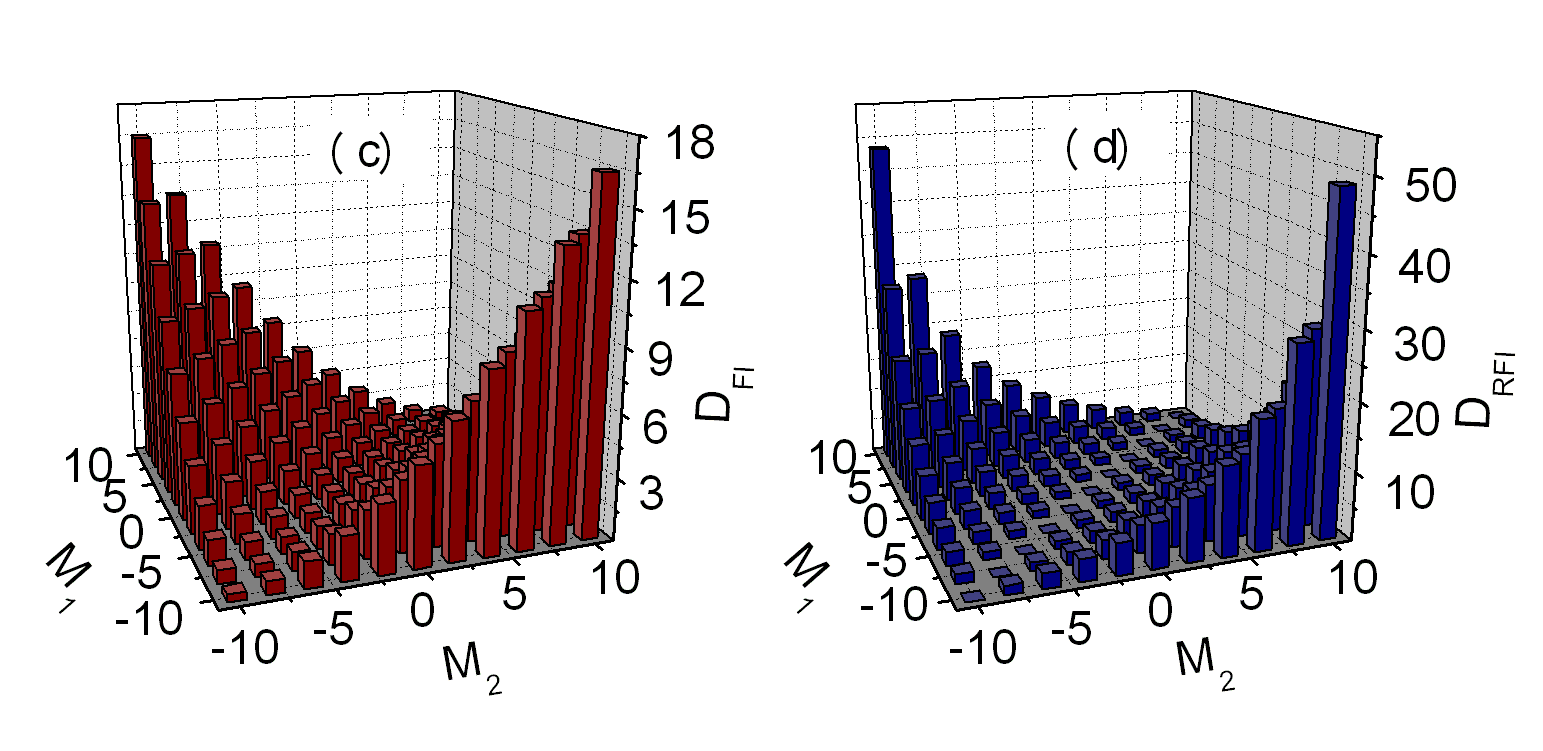}
\end{center}
\caption{(color online) (a, c) Size $D_{FI} ( \Psi ) $ of the linear superpositions $|\Psi\rangle$ formed by the ground states of the Mn$_{12}$ (a) and Fe$_8$ (c) molecules, as a function of the total-spin projections $M_1$ and $M_2$ of the two components. In the cases $M_1=M_2$ we set $D_{FI} =0$. 
(b, d) Size of the same linear superpositions based on the relative Fisher information, $D_{RFI} ( \Psi ) $, for Mn$_{12}$ (b) and Fe$_8$ (d). }
\label{Mn12FigA}
\end{figure}

We next quantify the size of the linear superpositions $ | \Psi \rangle $, where the components $ | \Psi_1 \rangle $ and $ | \Psi_2 \rangle $ are identified with arbitrary $ M_1 $ and $ M_2 $ ground states, by means of the quantum Fisher information. The dependence of the size $ D_{FI} ( \Psi ) $ on $M_1$ and $M_2$ is reported in Fig. \ref{Mn12FigA}(a). The largest value of $D_{FI} ( \Psi )$ is achieved by superimposing the maximally polarized components ($M_1 = -M_2 = 10$). Such value can be compared with the maximum in principle achievable within the present spin cluster, through a suitable engineering of the spin Hamiltonian, which is given by $ D_{FI} ( \Psi_{max} ) = 22 $.
Moreover, the size of the linear superposition shows a strong dependence on $|M_1-M_2|$, while it depends only weakly on $|M_1+M_2|$. At a qualitative level, the same features characterize the relative Fisher information $D_{RFI} (\Psi)$ (Fig. \ref{Mn12FigA}(b)). 

We next compute $ D_{LM} (\Psi , \delta = 10^{-2} )$ (Eq. \ref{eqlm}), in order to quantify the size of the linear superpositions $ | \Psi \rangle $ in terms of the distinguishability between the two components through local measurements. The probability $P_l$ of discriminating between $ | \Psi_1 \rangle $ and $ | \Psi_2 \rangle $ through single-spin measurements is reported in Fig. \ref{Mn12FigB}(a) for the three inequivalent spins, as a function of $M_1=-M_2$. In particular, we find that only one of the spins ($s^A_2$, green squares) reaches values of $P_l > 0.99 $, and only for $M_1=10$ \cite{indspinsMn12}. In order to estimate the size $D_{LM}$ for such case, we need to consider different partitions of the spin cluster, where the spins with least distinguishable states are grouped together in larger subsystems $A_l$, so as to achieve values of the probability $P_l$ above the threshold $1-\delta$. The partition with the largest number of subsystems that fulfils such requirement for $M_1=-M_2=10$ is defined by the four two-spin subsystems $ A_l = \{ s_{2l-1}^A,s_l^B) \} $ (purple squares), with $1 \le l \le 4$, and by four additional subsystems, each formed by an individual spin: $ A_{l+4} = \{ s_{2l}^A \} $. As a result, $D_{LM} (\Psi , \delta = 10^{-2} ) = 8$. 

The spin Hamiltonian we have considered so far is characterized by the presence of a dominant antiferromagnetic interaction ($J_{AB}$), and specifically that between the spins $s^A_i$ and $s^B_{2i-1}$ ($i=1, \dots , 4$). This tends to reduce the classical-like character of the polarized ground states ($M= \pm 10$), and the partial spin sums $S_A$ and $S_B$ corresponding to each sublattice. Both the features possibly limit the sizes of the linear superpositions between ground states (see also Subsec. \ref{subsec3}). In this respect, it's intructive to compare the above values of values of $D_{FI}$ and $D_{LM}$ with those obtained starting from another possible set of exchange couplings ($J_{AB}=67\,$K, $J_{AB}'=62\,$K, $J_A=6\,$K, and $J_B=8\,$K), which has been derived from high-energy inelastic neutron scattering \cite{Chaboussant04}. Here, the overlap between quantum and classical ground states is enhanced with respect to the previous value: $ | \langle M=10 | \Psi_{1}^{max} \rangle | = 0.589 $. Correspondingly, all the sizes $D_{FI} ( \Psi )$, $D_{RFI} ( \Psi )$, and $D_{LM} ( \Psi )$ of the linear superposition $ | \Psi \rangle $ between $M_1=10$ and $M_2=-10$ are to some extent increased (see Table \ref{tableFinal}).

\subsubsection{The Fe$_8$ molecule}  \label{subsec2}

The magnetic core of Fe$_8$ consists of eight Fe$^{3+}$ ions, each carrying an $s=5/2$ spin (Fig. \ref{Fe8Mn12}(a)). Experimental evidence exists that the Fe$_8$ molecule has a ground $S=10$ multiplet, resulting from a ferrimagnetic ordering of the spins \cite{Gatteschi}. The spin-Hamiltonian calculations allow the further specification of the spin ordering. This is characterized by six spins (sublattice $A$, blue circles) oriented parallel to each other and to the total spin, while the remaining two (sublattice $B$, red circles) are oriented in the opposite direction. The dominant part of the spin Hamiltonian includes the exchange couplings represented in the Fig. \ref{Fe8Mn12}(a), and is given by $ H_{Exc} = H_A + H_{AB} $, where:
\begin{eqnarray}
\label{HamFe8a}
H_A & = & J_A \sum_{i=1}^2 {\bf s}^A_{4+i} \cdot ( {\bf s}^A_{2i-1} + {\bf s}^A_{2i}) 
      +   J_A'             {\bf s}^A_5 \cdot {\bf s}^A_6 , \\
H_{AB} & = &  
J_{AB}  \sum_{i=5}^6 \sum_{j=1}^2 {\bf s}^A_{i} \cdot {\bf s}^B_{j} + \nonumber\\
\label{HamFe8b}
& & 
J_{AB}'
\left[
        {\bf s}^B_2 \cdot ( {\bf s}^A_{2} + {\bf s}^A_{3} ) 
+       {\bf s}^B_1 \cdot ( {\bf s}^A_{1} + {\bf s}^A_{4} ) \right] ,
\end{eqnarray} 
with $s_i^A=s_j^B=5/2$. A good agreement with the experimental data is obtained with the following set of values for the exchange constants: 
$J_A=26\,$K, $J_A'=36\,$K, $J_{AB}=200\,$K, $J_{AB}'=59\,$K \cite{Barra00,Carretta06,Raghu01}. 
As in the case of Mn$_{12}$, we take as a reference the linear superposition between $M_1=10$ and $M_2=-10$. The operator that maximizes $ \mathcal{F} ( \Psi )$ (Eq. \ref{explFI}) in such case is $ X = S_z^* $, corresponding to $ \hat{\bf n}_{i} = \hat{\bf z} $ ($ \hat{\bf n}_{i} = -\hat{\bf z} $) for the spins belonging to sublattice $A$ ($B$).

\begin{figure}[ptb]
\begin{center}
\includegraphics[width=8.0cm]{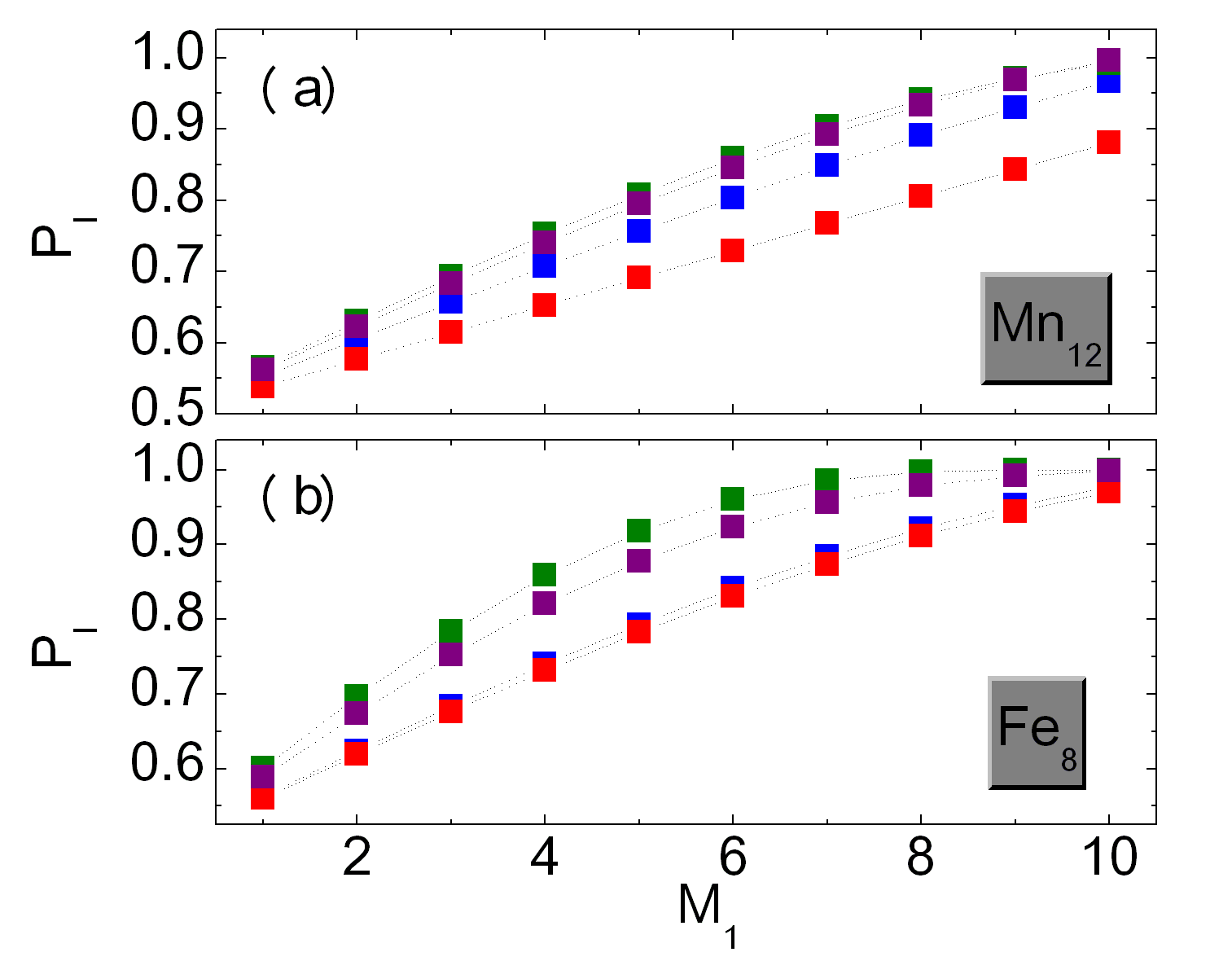}
\end{center}
\caption{(color online) Probability $P_l$ of discriminating between the two components $ | \Psi_1 \rangle $ and $ | \Psi_2 \rangle $ (with $M_1=-M_2=M$) of the Mn$_{12}$ (a) and Fe$_8$ (b) spin clusters. In the case of Mn$_{12}$, blue, green and red squares correspond to subsystems formed by individual spins: $s^A_1$, $s^A_2$ and $s^B_1$, respectively 
(Fig. \ref{Fe8Mn12}(b)). The purple squares correspond to subsystems formed by spin pairs such as ($s^A_1,s^B_1$). In the case of Fe$_8$, red, blue and green squares correspond to subsystems formed by individual spins: $s^A_5$, $s^B_1$ and $s^A_1$, respectively. The purple squares correspond to the subsystem formed by the four central spins $s^A_5$, $s^A_6$, $s^A_1$ and $s^A_2$ (Fig. \ref{Fe8Mn12}(a)).}
\label{Mn12FigB}
\end{figure}

We start by verifying to which extent the maximally polarized ground state of $H_{Exc}$ approaches the classical ground states (Eq. \ref{wfmax}). These represent the main spin configurations in the expression of $ | M=\pm 10 \rangle $, with an overlap given by $ | \langle M= 10 | \Psi_{1}^{max} \rangle | = 0.687 $ \cite{indspinsFe8}. The expectation value of the staggered magnetization is 
$ \langle M=10 | S_z^* | M=10 \rangle = 18.0 $ (to be compared with 
$ \langle \Psi_1^{max} | S_z^* | \Psi_1^{max} \rangle = 20.0 $),
while the variance is
$ \mathcal{V}_{M=10} (S_z^*) = 6.77 $.
On average, the ground state $ | M=10 \rangle $ thus resembles its classical counterpart slightly more in the Fe$_8$ spin cluster than in Mn$_{12}$.  
States with a lower $M$ are characterized by lower values and larger fluctuations of the staggered magnetization: the value of $\langle M | S_z^* | M \rangle$ is in fact proportional to $M>0$, whereas $\mathcal{V}_{M}$ decreases exponentially for increasing $M$. 

The sizes $ D_{FI} (\Psi) $ of the linear superpositions between pairs of ground states $| M_1 \rangle $ and $ | M_2 \rangle $ (Fig. \ref{Mn12FigA}(c)) are very similar to those obtained for the Mn$_{12}$ spin cluster (panel (a)). The same applies to the relative Fisher information, especially for the most relevant case: $ M_1=-M_2=\pm 10 $ (panel (d)). On the other hand, the both $ D_{FI} (\Psi) $ and $ D_{RFI} (\Psi) $ decrease faster than for Mn$_{12}$ with decreasing $ |M_1-M_2| $.
 
We finally consider the sizes $ D_{LM} ( \Psi , \delta = 10^{-2}) $ based on the distinguishability of the two components $ | \Psi_1 \rangle $ and $ | \Psi_2 \rangle $ by means of local measurements. The dependence of $ P_l $ on $ M_1 = -M_2 $ is reported in Fig. \ref{Mn12FigB}(b) for each of the three inequivalent spins in the cluster. Such probability is highest for the external spins belonging to the sublattice $A$ ($1 \le i \le 4$, green squares), and lies above the threshold $1-\delta$ for $M_1 \ge 9$. The probabilities corresponding to the central spins (red and blue squares) lie instead below the threshold for all values of $M_1$. In order to efficiently discriminate between the two components $| \Psi_k \rangle$, one needs to perform a measurement on the whole central core of the cluster (purple squares). Here, the condition $ P_5 > 0.99 $ is achieved for $ M_1 \ge 9 $. Therefore, $ D_{LM} ( \Psi , \delta = 10^{-2} ) = 5 $ for $M_1=9$ and $M_1=10$, where the optimum partition corresponds to $A_l = \{ s^A_l \} $ ($l=1,\dots ,4$) and $A_5=\{ s^B_{1} , s^B_{2} , s^A_{5} , s^A_{6} \}$. We can thus conclude that, while the smaller number of spins doesn't result in a smaller size $ D_{FI} ( \Psi ) $ for Fe$_8$ than for Mn$_{12}$, it does result in significantly smaller linear superpositions in terms of $ D_{LM} ( \Psi , \delta ) $ (Table \ref{tableFinal}), being the spin polarization of the former molecule concentrated in fewer spins. 

\subsection{Increasing the potential size of linear superpositions in high-spin nanomagnets} \label{subsec3}

Hereafter, we show how the size of the linear superposition $ | \Psi \rangle $ can be possibly increased with respect to the above values. In particular, we first consider the possibility of modifying the exchange couplings in the spin Hamiltonians of Mn$_{12}$ (Eqs. \ref{HamMn12a}-\ref{HamMn12b}) and Fe$_8$ (Eqs. \ref{HamFe8a}-\ref{HamFe8b}), and provide a general criterion for maximizing the sizes of $ | \Psi \rangle $ in spin clusters with a given geometry and a given ferrimagnetic ordering. Then, we discuss the dependence of such size on relative dimension of the two sublattices, with specific reference to a class of bipartite rings, which includes the Mn$_6$ molecule as a prototypical system, and to the star-shaped Fe$_4$ molecule \cite{Cornia04}. Finally, we discuss the limiting case represented by a spin clusters with only ferromagnetic interactions, such as the Mn$_{10}$ molecular magnet \cite{Caneschi97}. 

\subsubsection{''Improved'' Mn$_{12}$ and Fe$_8$ spin clusters}

The sizes of the linear superpositions that can be generated within the ground $S$ multiplets of Mn$_{12}$ and Fe$_8$ can be possibly increased by optimizing the exchange couplings (Eqs. \ref{HamMn12a}, \ref{HamMn12b}, \ref{HamFe8a}, \ref{HamFe8b}), without modifying the geometry of the spin clusters, nor their bipartition 
in the sublattices $A$ and $B$. 

As discussed in the previous Section, the theoretical maxima of the measures $D_{FI} (\Psi)$ and $D_{LM}(\Psi)$ are achieved by states of the form $ | \Psi_{max} \rangle $ (Eq. \ref{xxx}). However, except for the limiting case $N_B=0$, the components $ | \Psi_k^{max} \rangle $ in general cannot be obtained as maximally spin-polarized ground states of an exchange Hamiltonian. In fact, unlike such states, $ | \Psi_k^{max} \rangle $ doesn't correspond to an eigenstate of ${\bf S}^2$, and is characterized by the following relation between the total-spin and its projection along $z$:  
$ % \begin{equation}\label{eq99}
\langle \Psi_1^{max} | {\bf S}^2 | \Psi_1^{max} \rangle = 
M(M+1) + 2\sum_{j=1}^{N_B} s^B_j 
$, % \end{equation}
where $ M = \sum_{i=1}^{N_A} s^A_i - \sum_{j=1}^{N_B} s^B_j > 0 $.

Amongst the eigenstates of an exchange Hamiltonian, the states that maximize $ D_{FI} (\Psi) $, are those with maximum values of the partial spin sums, and with ${\bf S}_A$ and ${\bf S}_B$ antiparallel to each other:
\begin{equation} \label{yyy}
| \Psi_{k=1,2} \rangle \!=\! | S_A \!\!=\!\! S_A^{max} , S_B \!\!=\!\! S_B^{max} , M \!\!=\!\! \pm (S_A\!\!-\!\!S_B) \rangle ,  
\end{equation} 
where $ S_{\chi = A,B}^{max} = \sum_{i=1}^{N_\chi} s_i^\chi $. These correspond to the actual ground states of an exchange Hamiltonian in the limit of strong ferromagnetic coupling between spins of each sublattice, and weak antiferromagnetic interaction between spins of different sublattices: $J_A, J_B < 0$,  $J_{AB}>0$, and $ |J_A|, |J_B| \gg J_{AB} $. In the case of Mn$_{12}$, and given the sublattices represented in Fig. \ref{Fe8Mn12} (a), the maximum values of the partial spin sums are $S_A^{max}=16$ and $S_B^{max}=6$. 
The corresponding sizes are:
$
D_{FI} (\Psi) = 20.0
$,
$
D_{RFI} (\Psi) = 143.0
$,
and 
$
D_{LM} (\Psi) = 10
$.
For the Fe$_{ 8}$ spin cluster, instead, $S_A^{max}=15$ and $S_B^{max}=5$. 
This results in the following sizes of the linear superposition between the $M_1=10$ and $M_2=-10$ ground states:
$
D_{FI} (\Psi) = 18.3
$,
$
D_{RFI} (\Psi) = 151.9
$,
and
$
D_{LM} (\Psi) = 8
$.
Altogether, the amount of quantum fluctuations in the linear superposition (and thus $D_{FI} ( \Psi )$) is close to that obtained for realistic values of the exchange couplings, for both Mn$_{12}$ and Fe$_8$ (Table \ref{tableFinal}). On the other hand, ground states such as those reported in Eq. \ref{yyy} represent a better approximation of classical-like states, resulting in significantly increased values of the sizes $D_{RFI} ( \Psi )$ and $D_{LM} ( \Psi )$. 

\subsubsection{The Mn$_6$ and Fe$_4$ molecules}

Even in the absence of strong ferromagnetic interactions within each sublattice, the ground states of $H_{Exc}$ can approach the above form (Eq. \ref{yyy}) in systems with low spin frustration and highly inequivalent sublattices. Rings formed by alternate sequences of inequivalent spins represent a class of systems that can possibly meet such requirements. In particular, we consider the case where the spins $s_A$ and $ s_B < s_A $, with $N_A = N_B = N/2 $, are coupled to each other by isotropic exchange interaction between nearest neighbors (${\bf s}_{N/2+1}^B \equiv {\bf s}_1^B$):
\begin{equation} \label{mn6}
H_{AB} = J_{AB} \sum_{i=1}^{N/2} {\bf s}^A_i \cdot ( {\bf s}^B_i + {\bf s}^B_{i+1} )  ,
\end{equation} 
while no intra-sublattice interaction is present ($H_A=H_B=0$). The antiferromagnetic coupling ($J_{AB}>0$), combined with the spin difference $ s_A - s_B $, results in the formation an $ S = N ( s_A - s_B ) / 2 $ ground multiplet. A prototypical example of such a spin cluster is represented by the Mn$_6$ ring, formed six Mn$^{2+}$ ions ($s_A=5/2$) coupled to six organic radicals ($s_B=1/2$) \cite{Caneschi88}. In order to highlight the role played by the inequivalence between the two spins, this molecular nanomagnet will be compared with analogous spin rings, characterized by $N_A=N_B=6$, $s_B=1/2$ and $s_A=1,3/2,2$. We shall focus on the linear superposition that involves the maximally polarized ground states ($M_1=-M_2=S)$, for which the optimal X coincides with $S_z^*$.

\begin{figure}[ptb]
\begin{center}
\includegraphics[width=8.0cm]{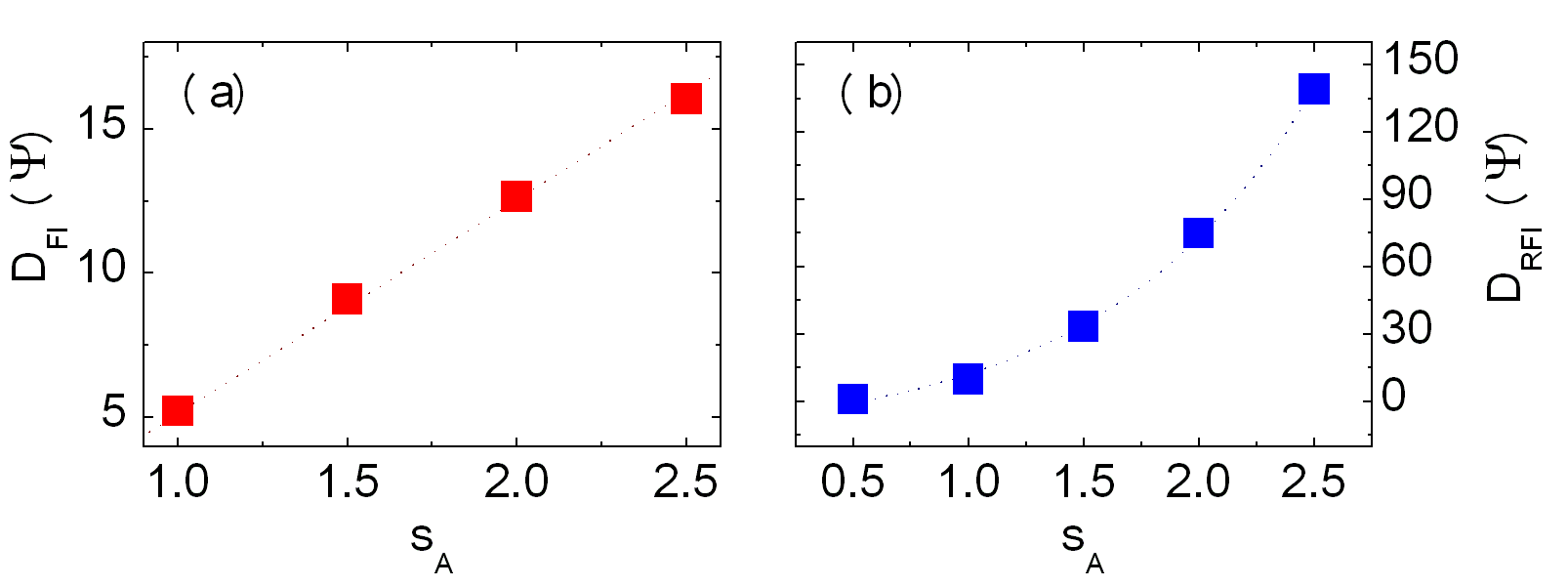}
\end{center}
\caption{(color online) (a) Size of the linear superposition of the $M_1=+S$ and $M_2=-S$ ground states of the bipartite rings, as a function of  $s_A$ (where $S = 6 (s_A-s_B) $ and $s_B=1/2$). The values of $D_{FI} (\Psi )$ (red squares) display a linear dependence on $s_B$ (dotted curve). 
(b) Size of the same linear superpositions, quantified by $D_{RFI} (\Psi )$. The dotted curve represents an exponential fit of the computed values (blue squares).}
\label{Mn6FigAB}
\end{figure}

The size of the linear superposition $ | \Psi \rangle $ clearly depends on the inequivalence between the two sublattices. In fact, the value of $D_{FI} (\Psi )$ (Fig. \ref{Mn6FigAB}(a)) grows linearly with $s_A$. The growth of the relative Fisher information (Fig. \ref{Mn6FigAB}(b)) is instead exponential, reflecting a fast decrease with $s_A$ of the quantum fluctuations in the ground states $ | \Psi_k \rangle $. The   largest value of $s_A$ we consider ($s_A=5/2$) corresponds to the case of the Mn$_6$ nanomagnet. This spin cluster is characterized by a classical-like ground state, and by average values of the partial spin sums $S_A$ and $S_B$ (14.7 and 2.68, respectively) that approach the theoretical maxima, $S_A^{max}=15$ and $S_B^{max}=3$. Also the size of the linear superposition approaches the value that would result from the state of Eq. \ref{yyy}, namely $D_{FI}=17.1$. Finally, the size based on the distinguishability between the components through local measurements is $D_{LM} (\Psi , \delta = 10^{-2}) = 7$. This corresponds to the partition of the spin cluster into the following subsystems: 
$A_l = \{ s^A_l \}$, with $l=1,\dots , 6$, and $ A_7 = \{ s^B_1 , s^B_2 , s^B_3 , s^B_4 , s^B_5 , s^B_6 \}$. 

As an alternative example of molecular spin cluster with highly inequivalent sublattices, we briefly mention the case of Fe$_4$ \cite{Cornia04}. The magnetic core of such molecule consists of three Fe$^{3+}$ ions, each carrying an $s^A_i=5/2$ spin, antiferromagnetically coupled to a fourth iron ion ($s^B_1=5/2$). The exchange term of the spin Hamiltonian reads:
\begin{equation} \label{fe4}
H_{AB} = J_{AB} \sum_{i=1}^{3} {\bf s}^A_i \cdot {\bf s}^B_1 ,
\end{equation} 
and its ground $S=5$ multiplet is characterized by a parallel alignment of the external Fe ions (sublattice $A$). 
The largest size corresponds to the linear superposition of the polarized ground states (see Table \ref{tableFinal}). These possess classical-like features 
($ | \langle M=5 | \Psi_1^{max} \rangle | = 0.829 $), 
and average values of the partial spin sum $S_A$ that coincides with the theoretical maximum ($S_A^{max}=15/2$). 
Correspondingly, the size of the linear superposition based on Fisher information reaches the value that would result from the state of Eq. \ref{yyy}, namely $D_{FI}=8.603$. Instead, the size based on the distinguishability between the components through local measurements is $D_{LM} (\Psi , \delta = 10^{-2}) = 3$. This corresponds to the partition of the spin cluster into the following subsystems: 
$A_l = \{ s^A_l \}$, with $ l = 1 , 2 $, and $ A_3 = \{ s^B_1 , s^A_3 \}$. Here, the external iron ions carry a large amount of which-component information, resulting in a probability $P_l > 0.99 $. The probability of discriminating between the two components based on the state of the internal ion is also large (0.937), but remains below such threshold.
Linear superpositions between the $M_1=5$ and $M_2=4$ ground states of Fe$_4$ have been recently generated by pulsed microwave fields \cite{Schlegel08}. The sizes of such linear superpositions, reported in Table \ref{tableFinal}, are comparable to those generated in the low-spin molecules Cr$_7$Ni and V$_{15}$ with the $S=1/2$ ground states. 

\subsubsection{Ferromagnetic spin clusters}

Molecular spin clusters with only ferromagnetic interactions can be regarded as the limiting cases of the ones considered so far, with $N_A=N$ and $N_B=0$. Here, the ground states with $M= \pm S = \pm \sum_{i=1}^N s_i$ coincide with $ | \Psi^{max}_{k=1,2} \rangle $, and the sizes of their linear superposition are given simply by 
$ D_{FI} ( \Psi ) = S $ and $ D_{LM} ( \Psi , \delta ) = N $. As a remarkable example of such a molecular nanomagnet, we mention [Et$_3$NH]$_2$[Mn(CH$_3$CN)$_4$(H$_2$O)$_2$][Mn$_{10}$O$_4$(biphen)$_4$Br$_{12}$], Mn$_{10}$ in short \cite{Caneschi97}. Such a spin cluster has an $S=23$ ground multiplet, and allows one in principle to generate the largest linear superpositions, amongst the considered molecules (Table \ref{tableFinal}). 
Single-atom nanomagnets, such as the lanthanide double-decker complexes \cite{Ishikawa03}, can be regarded as the limiting case of a ferromagnetic spin cluster. There, while the presence of a single magnetic ion implies $D_{LM} (\Psi , \delta) = 1 $ for all linear superpositions, the high value of the total momentum $ {\bf J} = {\bf L} + {\bf S} $ results in values of $D_{FI} (\Psi) = J$ which are comparable to those of large molecular spin clusters. In Table \ref{tableFinal}, we report as a representative example the case of the TbPc$_2$ single-atom nanomagnet.

For a given spin cluster, the polarized ground states of an exchange Hamiltonian with only ferromagnetic interactions clearly maximize all the measures discussed so far. One might notice however that the difficulty of generating a linear superposition (e.g., by means of microwave pulse sequences) increases with the difference between the components in terms of total-spin projection. 
For a given $S$, the theoretical maximum for the size of the linear superposition between the $M=\pm S$ ground states is given by $ D_{FI} ( \Psi_{max} ) = S + 2 S_B $, and is thus larger for ferrimagnetic ($S_B>0$) than for ferromagnetic ($S_B=0$) spin clusters. Therefore, the former systems might enable the generation of linear superpositions with larger sizes, if the constraint is on the value of $M_1-M_2$, rather than on the number ($N$) and values ($s_i$) of the spins. 

\section{Low-spin molecules}\label{SecLSM}

Molecular nanomagnets with dominant antiferromagnetic interaction are typically characterized by frustrated and low-spin ground states, resulting from the competition between different exchange interactions. In these systems it's not clear a priori to which extent quantum features, such as the fluctuations of single-spin operators, are actually enhanced by linearly superposing two ground states. Also, one might wonder to which extent a linear superposition between the $M = \pm 1/2$ ground states of an $S=1/2$ spin cluster differs from one that is generated with an individual $s=1/2$ spin. These issues are investigated in the present Section, by exploiting the measures based on quantum fluctuations of single-spin operators and state distinguishability through local measurements. We specifically refer to the Cr$_7$Ni \cite{Ardavan07} and V$_{15}$ \cite{Bertaina08} molecules, where linear superpositions have been experimentally demonstrated in recent years. 

\subsection{The Cr$_7$Ni molecule}

The magnetic core of the Cr$_7$Ni consists of seven Cr ions ($s=3/2$) and a Ni ($s=1$), spatially arranged so as to form a regular octagon. The dominant term in the spin Hamiltonian of the molecule is represented by the isotropic exchange between nearest neighbors \cite{Troiani05b} (Eq. \ref{mn6}, with $N=8$). Within each of the $S=1/2$ ground states, neighboring spins have antiparallel expectation values: one can thus define two sublattices $A$ and $B$, formed by even- and odd-numbered spins, respectively, with the latter ones including the Ni. Linear superpositions between the $M_1=1/2$ and $M_2=-1/2$ ground states have been experimentally generated by pulsed microwave fields \cite{Ardavan07}, and will be discussed hereafter. 

We preliminarily note that each component $ | \Psi_k \rangle $ is perfectly equivalent to the linear superposition $ | \Psi \rangle $ in terms of quantum fluctuations of single-spin operators. In fact, these states can be transformed into one another by an overall spatial rotation, that also transforms any combination $X$ of single-spin operators into another one. In particular, we find that:
\begin{equation}
\mathcal{V}_{\Psi  } (S_z^*) = 
\mathcal{V}_{\Psi_k} (S_x^*) = 45.836 ,
\end{equation}
out of a theoretical maximum of $ (\sum_{i=1}^8 s_i)^2 = (23/2)^2 = 132.25 $. Linearly superposing the two ground states $ | \Psi_k \rangle $ can however increase the quantum fluctuations of a given single-spin operator $X$. In particular, the particular $X$ that maximizes $ \mathcal{V}_\Psi $ and at the same time minimizes $ \mathcal{V}_{\Psi_k} $ is $ S_z^* = S_z^A - S_z^B $, being 
$ \langle s_{i,x} s_{j,x} \rangle = 
  \langle s_{i,y} s_{j,y} \rangle = 
  \langle s_{i,z} s_{j,z} \rangle $
for all the spin pairs, and 
$ \langle s_{i,\alpha} \rangle \neq 0 $ 
only for $\alpha = x$ ($\alpha = z$) in $ | \Psi \rangle $ ($ | \Psi_k \rangle $). 
This results in the following sizes of the linear superposition $ | \Psi \rangle $, based on the quantum Fisher information: 
\begin{equation}
D_{FI} (\Psi) = 3.986, \ \ D_{RFI} (\Psi) = 2.668 .
\end{equation}
The former measure can be contrasted on the one hand with the theoretical maximum for the present spin cluster, that is $ \sum_{i=1}^8 s_i = 11.5$, and on the other hand with the size of the same linear superposition generated with an individual $s=1/2$ spin, for which $D_{FI} (\Psi) = 1/2$. Such comparisons show that the linear superposition between the ground states of Cr$_7$Ni is quite smaller with respect to those can be possibly generated in high-spin nanomagnets, and remains farther from its theoretical maximum (Table \ref{tableFinal}). However, the size of the state $ | \Psi \rangle $ in this spin cluster is significantly larger than that of an individual $s=1/2$ spin \cite{rinota}. 

An analogous conclusion can be drawn from the distinguishability of the two components $|\Psi_k\rangle$ through local measurements. The maximum number of subsystems into which the ring can be divided such that each subsystem carries the which-component information is in fact:
\begin{equation}
\mathcal{D}_{LM} ( \Psi ) = 2 . 
\end{equation}
In particular, any bipartition of the ring into groups of four spins allows one to achieve values of the probability $ P_l > 0.99 $, whereas this can be done with none of the partitions into a larger number of subsystems. 

\begin{table*}
\begin{tabular}{c|ccccccccccc}
\hline \hline
Molecule                              & \ Mn$_{12}(1)$\ & \ Mn$_{12}(2)$\ & \ \ Fe$_{8}$\ \ & \ \ Mn$_{6}$\ \ & \ \ Mn$_{10}$\ \ & \ \ Tb\ \ & \ Fe$_4(1)$\ & \ Fe$_4(2)$\ & \ \ Cr$_7$Ni\ \ & \ V$_{15}(1)$\ & \ V$_{15}(2)$\ \\
\hline 
$M_1-M_2$                             & 20    & 20    & 20    & 24    & 46   & 12 & 10   & 
 1 & $ 1  $ & $ 1   $ & $ 3  $ \\
$ D_{ FI} (\Psi                   ) $ & 14.4  & 19.3  & 16.5  & 16.0  & 23.0 & 6  & 8.603& 
 0.366 & 3.986 & 1.478 & 1.544 \\
$ D_{ FI} (\Psi_k                 ) $ & 0.318 & 0.170 & 0.339 & 0.115 & 0.0  & 0  & 0.200& 
 0.282 & 1.494 & 1.361 & 1.244 \\
$ D_{RFI} (\Psi                   ) $ & 45.4  & 113.0 & 48.7  & 139   & $-$  &$-$ & 43.07& 
 1.299 & 2.668 & 1.086 & 1.241 \\
$ D_{LM} (\Psi , \delta = 10^{-2})  $ & 8     &  9    & 5     & 7     & 10   & 1  & 3     & 
 1     & 2     & 1     & 3 \\
\hline \hline
\end{tabular}
\caption{\label{tableFinal} Size of the linear superposition $ | \Psi \rangle $ for different molecular nanomagnets. The molecules Cr$_7$Ni and V$_{15}$ are discussed in Sec. \ref{SecLSM}, all the others in Sec. \ref{SecHSM}. In the case of Fe$_4(2)$, we consider the linear superposition between the $S=5$ ground states with $M_1=5$ and $M_2=4$, generated in Ref. \cite{Schlegel08}. The column corresponding to V$_{15}(2)$ refers to a linear superposition between the $M_1=3/2$ and $M_2=-3/2$ eigenstates of the  $S=3/2$ quadruplet. In all the other cases, the two components $ | \Psi_k \rangle $ coincide with the states $M_1=S$ and $M_2=-S$ of the ground $S$ multiplet. For the Mn$_{12}$ molecular nanomagnet, we have considered both the set of exchange parameters suggested in Ref. \cite{Raghu01} (Mn$_{12}$(1)) and that derived in Ref. \cite{Chaboussant04} (Mn$_{12}$(2)). As a single-ion magnet, we consider the Tb double-decker complex, with a $J=6$ ground state \cite{Ishikawa03}.} 
\end{table*}

\subsection{The V$_{15}$ molecule}

The V$_{15}$ spin cluster consists of $N=15$ oxovanadium ions, each carrying an $s=1/2$ spin \cite{Bertaina08}. The spins are arranged in three layers so as to form a triangle ($T$), sandwiched between two hexagons ($H1$ and $H2$). The spins that form each hexagon are coupled by a strong antiferromagnetic interaction, and are thus approximately frozen in a singlet state ($S_{H1}=S_{H2}=0$) at low temperatures. Besides,  they mediate an effective interaction between the spins belonging to the triangle, described by the following spin Hamiltonian: 
\begin{equation}
H_T = J \sum_{i=1}^3 {\bf s}^T_i \cdot {\bf s}^T_{i+1} 
+ D \hat{\bf z} \cdot \sum_{i=1}^3 {\bf s}^T_i \times {\bf s}^T_{i+1} ,
\end{equation}
where ${\bf s}_4 \equiv {\bf s}_1$.
In this effective three-spin model, the two states involved in the linear superposition $ | \Psi \rangle $ differ only with respect to the component of the spin triangle:
\begin{equation}\label{v15}
| \Psi_{k=1,2} \rangle = | \Psi^{H1} \rangle \otimes | \Psi^T_k \rangle \otimes | \Psi^{H2} \rangle ,
\end{equation}
where $ | \Psi^{H1} \rangle $ and $ | \Psi^{H2} \rangle $ both correspond to the singlet ground state of the hexagons, and $ | \Psi_{k=1,2}^T \rangle $ are different eigenstates of $H_T$. 
As a consequence, the expectation values of $X$ and $X^2$ take the following simplified forms:
\begin{eqnarray}
\langle \Psi_k | X | \Psi_k \rangle \!\!\! & = & \!\!
\langle \Psi_k^T | X_T | \Psi_k^T \rangle
\\
\langle \Psi_k | X^2 | \Psi_k \rangle \!\!\! & = & \!\!
\langle \Psi_k^T | X^2_T | \Psi_k^T \rangle
+ 2 
\langle \Psi_{H1} | X^2_{H1} | \Psi_{H1} \rangle ,
\end{eqnarray}
where the last expectation value accounts for the contribution from each of the two hexagons, and $ X_{\chi} = \sum_{i} \hat{\bf n}_{i}^{\chi} \cdot {\bf s}_{i}^{\chi} $ (with $\chi = H1,H2, T$).
 
We first consider the linear superposition involving the two $S_T=1/2$ eigenstates of the spin triangle with $+1$ eigenvalue of the spin chirality $ C_z = (4/\sqrt{3})\, {\bf s}_1^T \cdot {\bf s}_2^T \times {\bf s}_3^T $, that are coupled by a magnetic-dipole transition: 
\begin{eqnarray}
  | \Psi_1^T \rangle & \!\!\!=\!\!\! & (               | \uparrow  \downarrow\downarrow \rangle \!+\!
                             e^{ i2\pi /3} | \downarrow\uparrow  \downarrow \rangle \!+\!
                             e^{-i2\pi /3} | \downarrow\downarrow\uparrow   \rangle ) 
/ \sqrt{3}
\nonumber\\
  | \Psi_2^T \rangle & \!\!\!=\!\!\! & (               | \downarrow\uparrow  \uparrow   \rangle \!+\!
                             e^{ i2\pi /3} | \uparrow  \downarrow\uparrow   \rangle \!+\!
                             e^{-i2\pi /3} | \uparrow  \uparrow  \downarrow \rangle ) 
/ \sqrt{3} .
\end{eqnarray}
In this case, the fluctuations of $X$ for the state $ | \Psi \rangle $ are maximized by 
$ \hat{\bf n}^T_i = \cos (2\pi i / 3) \hat{\bf y} + \sin (2\pi i / 3) \hat{\bf z} $ 
and by
$ \hat{\bf n}^{H1}_i = \hat{\bf n}^{H2}_i = (-1)^i \hat{\bf z} $.
The corresponding value of the variance, whose theoretical maximum is $ ( \sum_{i=1}^{15} s_i )^2 = (15/2)^2 = 56.25 $, is given by
\begin{equation}\label{v15p}
\mathcal{V}_\Psi (X) = 7/4 + 2 \times 4.6641 = 11.08 ,
\end{equation}
where the former contribution comes from the triangle and the latter one from the two hexagons \cite{notaV15}. This results in the following sizes of the linear superposition $ | \Psi \rangle $:
\begin{equation}
D_{FI} (\Psi) = 1.478, \ \ D_{RFI} (\Psi) = 1.086 .
\end{equation}

The two components $ | \Psi_1 \rangle $ and $ | \Psi_2 \rangle $ are not efficiently distinguishable by means of local measurements. In particular, the two hexagons carry no which-component information (Eq. \ref{v15}), and the two states $| \Psi^T_k \rangle$ of the spin triangle can only be distinguished with probability $P_l>0.99$ by considering all three spins. The size of linear superposition $ | \Psi \rangle $ based on local measurements thus takes the same value that it would have for an individual $s=1/2$ spin:
\begin{equation}
\mathcal{D}_{LM} ( \Psi ) = 1 .
\end{equation}

Linear superpositions in the V$_{15}$ molecule have also been generated in the $S_T=3/2$ excited quadruplet \cite{Bertaina08}. Hereafter we consider the case where the components coincide with the two maximally polarized eigenstates of $H_T$, namely 
\begin{eqnarray}
  | \Psi_1^T \rangle = | \uparrow  \uparrow  \uparrow   \rangle ,
\ \
  | \Psi_2^T \rangle = | \downarrow\downarrow\downarrow \rangle .
\end{eqnarray}
It is easily seen that the optimal operator $X_T$ coincides in this case with the spin projection along $z$ ($ \hat{\bf n}^T_i = \hat{\bf z} $), while the expressions of the optimized $X_{H1}$ and $X_{H2}$ coincide with those reported above. Due to the enhanced contribution from the spin triangle, the variance of $X$ increases to 
$ \mathcal{V}_{\Psi} = 11.58$. Correspondingly, the sizes of the linear superposition based on the Fisher information become:
\begin{equation}
D_{FI} (\Psi) = 1.544, \ \ D_{RFI} (\Psi) = 1.241 .
\end{equation}
Finally, being the two states $| \Psi^T_k \rangle$ fully distinguishable at the single-spin level, the size of $ | \Psi \rangle $ based on local measurement is:
\begin{equation}
\mathcal{D}_{LM} ( \Psi ) = 3 .
\end{equation}
Altogether, in spite of the large number of spins that form the V$_{15}$ cluster, the sizes of the linear superpositions that can be generated within its lowest multiplets are either similar ($D_{FI}$) or identical ($D_{LM}$) to those of a simple spin triangle. This is due to the particular form of the low-energy eigenstates (Eq. \ref{v15}), where each hexagon is approximately frozen in a singlet state. The hexagons thus provide a sizeable contribution to the size based on the overall quantum fluctuations in the state $ | \Psi \rangle $, but not to the ones that depend on the quantum features resulting specifically from the linear superposition.

\section{Summary and conclusions}

Resorting to different quantum-information theoretic measures, we have quantitatively investigated the size of linear superpositions that can be generated in different molecular nanomagnets (Table \ref{tableFinal}). 

Amongst the high-spin systems, the sizes obtained for the spin-polarized ground states of Mn$_{12}$ and Fe$_{18}$ are of the order of the spin number $N$ \cite{notaMisure}, and thus comparable to those obtained in mesoscopic system \cite{Marquardt08,Korsbakken10}. 
The considered measures can be maximized in ferrimagnetic systems by tuning the exchange couplings, such that the partial spin sum corresponding to each sublattice is maximal. This feature characterizes the ground state of an exchange Hamiltonian in either of the two following cases: $(i)$ in an arbitrary geometry, if the coupling between two spins is either strongly ferromagnetic or weakly antiferromagnetic, depending on whether the two spins belong to the same sublattice or to different ones; $(ii)$ in the presence of antiferromagnetic couplings only, if the spin cluster is bipartite and the two sublattices are highly inequivalent (as in the case of Mn$_6$). 

The systems where the size of the linear superpositions are maximum, given the number and values of the constituent spins, are represented by those with ferromagnetic exchange interactions only (Mn$_{10}$).  On the other hand, ferrimagnetic systems might allow the achievement of larger sizes 
if the constraint is on the value of $M_1-M_2$ instead (consider, for example, the comparison between the cases of low-spin molecules and Fe$_4$ in Table \ref{tableFinal}, all with $M_1-M_2=1$). We note that the difficulty of generating a linear superposition between ground states increases with the difference between their spin projections. The latter constraint might thus in practice more relevant than the former one.

In the case of low-spin nanomagnets, we address the question of to which extent the size of linear superposition is enhanced by the composite character of the molecule spin. Here, we have considered two molecular spin clusters, namely Cr$_7$Ni and V$_{15}$, characterized both by an $S=1/2$ ground state, but substantially different in terms of geometry, exchange pattern, and spin correlations. 
In both cases, the size of the linear superpositions between ground states is larger than those achievable with an individual $s=1/2$ spin, but not proportionate to the number and value of the constituent spins. In particular,  the ground states of V$_{15}$ are poorly distinguishable by means of local measurements, due to the presence of spin subsensembles whose state is approximately frozen in the low-energy multiplets of the spin Hamiltonian.  

The measures considered in the present paper can be usefully complemented by those based on different criteria, such as the fragility of the linear superposition with respect to decoherence \cite{Dur02}. In particular, the correlation between such sizes and those derived from the fluctuations of single-spin operators or from local measurements might be significant if the environment couples locally to the electron spins. This seems to be the case with the nuclear-spin bath, which discriminates between $ | \Psi_1 \rangle $ and $ | \Psi_2 \rangle $ on the basis of the single-spin expectation values corresponding to the two components, rather than only on $M_1-M_2$ \cite{Troiani12}. A detailed understanding of the different decoherence mechanisms \cite{Morello07,Takahashi11,Leuenberger00} represents a preliminary requirement for exploring such connections further, and for generating macroscopic and yet robust linear superpositions in molecular nanomagnets \cite{Frowis11}.

We finally comment on the possibility of estimating the size of linear superpositions in molecular nanomagnets by experimental means, and specifically by static measurements.
The fluctuations of the $X$ operator can always be written as combinations of spin-pair correlation functions, which can be selectively accessed by inelastic neutron scattering. Through this technique, one can thus quantify the size of a linear superposition $|\Psi>$, provided that this corresponds to the nondegenerate ground state of the system, and that the temperature is lower than the energy gap 
between ground and first-excited states. In the particular case of ferromagnetic systems ($N_B=0$), the operator $X$ coincides with $S_z$, and its fluctuations with the $z$ component of the magnetic susceptibility.
The measure based on the distinguishability between two components can be indirectly estimated by chemically-selective techniques, such as nuclear magnetic resonance or X-ray magnetic circular dichroism. These techniques give access to single-spin expectation values \cite{Arrio99,Ghirri09,Micotti06}, that can be used to derive lower bounds for the which-component information carried by each spin. Here, an experimental evidence of the phase coherence between the two components should be provided by different means.

\section*{Acknowledgements}

This work has been financially support by the FIRB project RBFR12RPD1 of the Italian MIUR, by the ARO MURI grant W911NF-11-1-0268, and by NSF grant numbers PHY-969969 and PHY-803304.
Useful input from  J.I. Cirac is gratefully acknowledged.  

% \bibliography{paper}

\begin{thebibliography}{50}
\bibitem[{ind({\natexlab{a}})}]{indspinsMn12}
\bibinfo{note}{The external spins (sublattice $A$) present the highest degree
  of polarization in the ground state $M_1=10$, especially the even-numbered
  ones $s_{2i}^A$. This is also reflected by the occupations of the single-spin
  states with $|m_i| = s_i$: $ p ( m_{1}^A = 2 ) = 0.692 $, $ p ( m_{2}^A = 2 )
  = 0.845 $, and $ p ( m_{1}^B = - 3/2 ) = 0.591 $.}
\bibitem[{ind({\natexlab{b}})}]{indspinsFe8}
\bibinfo{note}{The single-spin states corresponding to the maximally polarized
  eigenstate $ M_1 = 10 $ are characterized themselves by a strong degree of
  polarization, where the largest populations correspond to the states with the
  largest spin projection: $ p ( m_{1}^A = 5/2 ) = 0.976 $, $ p ( m_{5}^A = 5/2
  ) = 0.668 $, and $ p ( m_{1}^B = - 5/2 ) = 0.643 $. Remarkable differences
  emerge in this respect between the four external spins ($\chi =A$, with $1\le
  i\le 4$) and the four central ones ($\chi =A$, with $5\le i\le 6$, and $\chi
  = B$).}
\bibitem[{not({\natexlab{a}})}]{notaV15}
\bibinfo{note}{As in the case of the Cr$_7$Ni molecule, the same amount of
  quantum fluctuations can be obtained for each of the components $ | \Psi_k
  \rangle $ for a suitable operator $X$. In particular, $\mathcal{V}_{\Psi_k}
  (X) $ are maximized by set versors $ \hat{\bf n}^T_i = \cos (2\pi i / 3)
  \hat{\bf x} + \sin (2\pi i / 3) \hat{\bf y} $ and $ \hat{\bf n}^{H1}_i =
  \hat{\bf n}^{H2}_i = (-1)^i \hat{\bf z} $.}
\bibitem[{not({\natexlab{b}})}]{notaMisure}
\bibinfo{note}{In order to use the Fisher information for comparing the present
  sizes with those obtained in other systems, the measure $D_{FI} (\Psi)$
  should be divided by the average spin value $\sum_{i=1}^N s_i / N$. Such
  further normalization compensates for the fact that the single-particle
  operators entering the expression of $X$ don't have a unit operator norm, as
  in the original definition of the measure \cite{Frowis12}.}
\end{thebibliography}

\end{document}